\documentclass[12pt]{article}
\usepackage{amssymb}
\usepackage{amsfonts}
\usepackage{amsmath}
\usepackage{amsmath,amsthm}
\usepackage{graphicx,epsfig}
\RequirePackage[numbers,sort&compress]{natbib}

\begin{document}
\begin{center}
\Large{\bf Thermodynamics in $f(T)$ gravity with nonminimal coupling to matter}\\
\small \vspace{1cm} {\bf Tahereh Azizi\footnote{t.azizi@umz.ac.ir}}\quad and
\quad {\bf Najibeh Borhani\footnote{n.borhani@umz.ac.ir}}\\
\vspace{0.5cm} {\it Department of Physics, Faculty of Basic
Sciences,\\
University of Mazandaran,\\
P. O. Box 47416-95447, Babolsar, Iran}
\end{center} \vspace{1cm}

\begin{abstract}
In the present paper, we study the thermodynamics behavior of the field equations for the generalized $f(T)$
gravity with an arbitrary coupling between matter and the torsion scalar.
 In this regard, we explore the verification of the first law
 of thermodynamics at the apparent horizon of the Friedmann-Robertson-Walker universe in two different perspectives
 namely the non-equilibrium and equilibrium descriptions of thermodynamics. Furthermore, we investigate the validity of the
 second law of thermodynamics for both descriptions of this scenario with assumption that the temperature of matter
inside the horizon is similar to that of horizon.
\end{abstract}

\section{Introduction}
\label{intro}
The teleparallel equivalent of General Relativity
(TEGR) \cite{1,2} is an equivalent formulation of classical
gravity, in which instead of using the curvature defined via
the Levi-Civita connection, uses the Weitzenb\"{o}ck connection
that has no curvature but only torsion. This approach
 is closely related to the standard General
Relativity, differing only in "boundary terms" involving
total derivatives in the action.
 In this setup, the dynamical objects are the four linearly independent
vierbeins and the Lagrangian density, $T$, is constructed from the torsion
tensor which is formed solely from products of
the first derivatives of the vierbein \cite{2}.
However, in a similar manner to the $f(R)$ modified gravity
 \cite{Capozziello03,Nojiri06,Nojiri07,Sotiriou08,Nozari09},
the teleparallel gravity is generalized
to a modified $f(T)$ version \cite{Ferraro07,Bengochea09,Linder10,Hamani11} which the Lagrangian
density is an arbitrary function of the torsion scalar $T$.
 This modification enable the theory to explain
the late time acceleration of the universe \cite{Bengochea09,Bamba11a,Dent11} which
is favored by the observational data. So there is no need to introduce a
 mysterious dark energy component
for the matter content of the universe. The significant
advantage of the $f(T)$ gravity is that, the field equations are
second order differential equations and are
more manageable compared to the $f(R)$ theories.
 For some gravitational and cosmological aspects of the
modified teleparallel gravity see \cite{cai14}.
 Recently, a further generalization of the teleparallel
gravity has been introduced in \cite{Harko14a}
by considering a nonminimal coupling between matter
and the torsion scalar in the action.
 In this model, the gravitational field can be described in
terms of two arbitrary functions of the torsion scalar $T$,
namely $f_1(T)$ and $f_2(T)$, with the function $f_2(T)$ linearly
coupled to the matter Lagrangian \cite{Harko14a}. This nonminimal
torsion-matter coupling scenario can offer a unified description of the universe evolution,
from its inflationary to the late-time accelerated phases \cite{Harko14a}.
In \cite{Feng15} the energy conditions of this model is studied and
the validity of energy bounds is examined. The dynamical system analysis for the cosmological applications of this
model is carried out in \cite{Carloni16}.

 In the present work we are going to study the thermodynamics aspects of this
nonminimally coupled $f(T)$ model
at the apparent horizon of an expanding cosmological background.
Indeed, the black hole thermodynamics, set up
connections between general relativity
and the laws of thermodynamics \cite{Haw75}. In this content,
 a temperature and entropy,
which are proportional to the surface gravity and area of
the horizon respectively, are associated with the black hole.
The first law of black hole thermodynamics is given by the
identity $TdS=dM$ \cite{Bek73} where $M$ is the mass of the black hole.
 Furthermore, Jacobson \cite{Jac95} showed that the Einstein's
equations can be derived from the fundamental relation $dQ=TdS$
in all local Rindler horizons where $\delta Q$ and $T$ are the
energy flux across the horizon and Unruh temperature, respectively.
This approach soon generalized to the cosmological situation where it was shown that
by applying the Clausius relation to the apparent horizon of the Friedmann-Robertson-Walker
(FRW) universe, the Friedmann equation can be rewritten in the
form of the first law of thermodynamics \cite{Cai05}.
Recently, the equivalence of the Clausius relation
and the gravitational field equations has been investigated
to the more general
modified theories of gravity such as Gauss-Bonnet gravity
\cite{akbar07}, Lovelock gravity \cite{cai07,cai08}, Braneworld
gravity \cite{Sheykhi07}, scalar-tensor gravity \cite{Wu08}, $f(R)$ theories
\cite{Eling06,Bamba10a,Bamba10b,Bamba11b,Karami12a} and the extended models of $f(R)$-gravity
\cite{Bamba10c,Sharif12a,Harko14b,Sharif12b,Sharif15,Azizi15}. In the context of $f(T)$ gravity,
the first law of black hole
thermodynamics has been studied in \cite{Miao11} and the thermodynamics of the apparent
 horizon of the FRW universe is explored in \cite{Karami12b,Geng11,Sharif13a}.
 This issue is also studied in some modified $f(T)$ scenarios \cite{Zubair15a,Zubair15b,Zubair15c,Askin15}.

 On the other hand, in \cite{Eling06} it was pointed that in order to derive the field equations of $f(R)$
modified gravity, one should employ a non-equilibrium thermodynamics treatment. However,
it has been demonstrated in \cite{Bamba10b} that it is possible to obtain an equilibrium description of
thermodynamics on the apparent horizon of $f(R)$ gravity.
The same works also has been carried out in $f(T)$ gravity \cite{Geng11}
and some extended models of $f(R)$ gravity \cite{Bamba10b,Azizi15}.
In addition to the first law of thermodynamics, there have been a lot
of interest on exploring the second low of thermodynamics in gravitational theories
\cite{Dav87,Sad06,Set06,Bab05,Zho07,Izq06,Jam11,Abd14,Wan08}.
 According to the second law  of thermodynamics, the sum
of the horizon entropy and the entropy of the matter field, i.e. the
total entropy, is a non-decreasing function of time.

In this paper, we explore the laws of thermodynamics in
both non-equilibrium and equilibrium descriptions in the nonminimal $f(T)$ gravity model.
The organization of the paper is as follows: In section 2,
we briefly review the nonminimally
torsion-matter coupling model and its equations of motion.
In section 3, we treat a non-equilibrium descriptions of
thermodynamics and investigate the first and
second laws of thermodynamics.
We explore the equilibrium description of thermodynamics in section 4.
Finally, our conclusion will be appeared in section 5.

\section{The equations of motion}
\label{sec:1}
In the context of the teleparallel gravity, the dynamical object is
a vierbein field $e_i(x^{\mu})$, $i=0,1,2,3$,
which is an orthonormal basis for the tangent space at each point $x^{\mu}$ of
the manifold. The metric tensor is obtained
from the dual vierbein as $g_{\mu\nu}(x)=\eta_{ij}e_{\mu}^{i}(x)e_{\nu}^{j}(x)$
 where $\eta_{ij}=e_{i}.e_{j}$ is the Minkowski metric and
$e^{\mu}_{i}$ is the component of the vector $e_i$ in a coordinate basis.
 Note that the Greek indices label coordinates on the
manifold while Latin indices refer to the tangent space. The torsion tensor is defined as
\begin{equation}
T^{\lambda}_{~\mu\nu}=e^{\lambda}_{i}(\partial_{\mu}e_{\nu}^{i}-\partial_{\nu}e_{\mu}^{i})\,.\label{tele1}
\end{equation}
Defining other two tensors:
\begin{equation}
S^{~\mu\nu}_{\lambda}\equiv\frac{1}{2}(K^{\mu\nu}_{~\lambda}+\delta_{\lambda}^{\mu}T^{\sigma\nu}_{~\sigma}-
\delta_{\lambda}^{\nu}T^{\sigma\mu}_{~\sigma})\,\label{tele2}
\end{equation}
and
\begin{equation}
K^{\mu\nu}_{~\lambda}\equiv-\frac{1}{2}(T^{\mu\nu}_
{~\lambda}-T^{\nu\mu}_{~\lambda}-T^{~\mu\nu}_{\lambda})\,,\label{tele3}
\end{equation}
one can write down the torsion scalar $T\equiv S^{~\mu\nu}_{\lambda}T_{~\nu\mu}^{\lambda}$. Using the
torsion scalar as the teleparallel Lagrangian leads to the same gravitational equations of the
general relativity. In this work, we focus on the modified teleparallel gravity with
a nonminimal coupling between the torsion scalar and the matter Lagrangian which is
introduced via the following action \cite{Harko14a}
\begin{equation}
S=\frac{1}{16\pi G}\int d^{4}x|e|\left\{T+f_{1}(T)+\left[1+\lambda f_{2}(T)
\right]\mathcal{L}_{m}\right\}\label{action}
\end{equation}
where $e=det(e_{\mu}^{i})=\sqrt{-g}$, $f_{1}(T)$ and $f_{2}(T)$ are
arbitrary functions of the torsion scalar
and $\lambda$ is a coupling constant with units
of mass$^{-2}$. Varying the action with respect to the vierbein leads to the field equations \cite{Harko14a}
\begin{equation}
\begin{split}
&(1+F_{1}+\lambda F_{2}\mathcal{L}_{m})[e^{-1}\partial_{\mu}(ee^{\alpha}_{A}
S^{~\rho\mu}_{\alpha})-e^{\alpha}_{A}T^{\mu}
_{~\nu\alpha}S^{\nu\rho}_{\mu}]+(F'_{1}+\lambda F'_{2}\mathcal{L}_{m})
\partial_{\mu}Te^{\alpha}_{A}S^{~\rho\mu}_{\alpha}
+\frac{1}{4}e^{\rho}_{A}(f_{1}+T)\\
&-\frac{1}{4}\lambda F_{2}\partial_{\mu}
Te^{\alpha}_{A}\overset{\mathbf{em}}{S}_{\alpha}{}^{\rho \mu}+ \lambda F_{2}\,
e^{\alpha}_{A} S_{\alpha}{}^{\rho\mu} \, \partial_{\mu}\mathcal{L}_{m}=4\pi G \left(1+\lambda f_{2}\right) e^{\alpha}_{A}
\overset{\mathbf{em}}{T}_{\alpha}{}^{\rho},
\label{field1}
\end{split}
\end{equation}
where $F_{i}=df_{i}/dT$ and the prime denotes a derivative with respect to the torsion scalar
and we have defined $\overset{\mathbf{em}}{S}_{A}{}^{\,\rho
\mu}=\frac{\partial\mathcal{L}_{m}}{\partial{\partial_{\mu}{e^{A}_{\rho}}}}$.
 We assume that the matter content of the universe is given by a perfect fluid and the matter
Lagrangian density is described by
${{\cal{L}}_{m}}=-\rho_{m}$ which leads to $\overset{\mathbf{em}}{S}_{A}^{\,\rho\mu}=0$.
 So the energy momentum tensor of the matter is given by
\begin{equation}
\overset{\mathbf{em}}{T}_{\mu\nu}=(\rho_{m}+p_{m})u_{\mu}u_{\nu}-p_{m}g_{\mu\nu},\label{T(M)}
\end{equation}
where $u^{\mu}$ is the four velocities of the fluid in the comoving coordinates.
 For a flat homogeneous and isotropic Friedman-Robertson-Walker (FRW) universe, the vierbein is given by
\begin{equation}
e^{i}_{\mu}=diag[1,a(t),a(t),a(t)],\label{ vierbein}
\end{equation}
where $a(t)$ is the cosmological scale factor. Using the above
relation together with (\ref{tele1}), and (\ref{tele2}), one obtains $T=-6H^{2}$ where
$H=\frac{\dot{a}}{a}$ is the Hubble parameter.
The substitution of the FRW vierbein (\ref{ vierbein}) in the field equation
(\ref{field1}) yields the modified Friedmann Equations as follows
\begin{equation}
H^{2}=\frac{1}{3\mathcal{F}}\left[8\pi G\left(1+\lambda f_{2}\right)
\rho_{m}-\frac{f_{1}}{2}\right],\label{Friedmann1}
\end{equation}

\begin{equation}
 \dot{H}=-\frac{1}{\mathcal{F}}\bigg[4\pi G\left(\rho_{m}+P_{m}\right)
\left[1+\lambda(f_{2}-2TF_{2})\right]+H(\dot{F}_{1}-16\pi G\lambda\rho_{m}\dot{F}_{2})\bigg]\label{Friedmann2}
\end{equation}
where $\dot{F}_{i}=dF_{i}/dt$ and
$\mathcal{F}=\left(1+2F_{1}-32\pi G\lambda\rho_{m}F_{2}\right)$.
 Note that the usual $f(T)$ gravity can be recovered in the limit $\lambda = 0$.
It has been shown that equations (\ref{Friedmann1}) and (\ref{Friedmann2}) can describe the acceleration expansion
of the universe without introduction of any dark energy component \cite{Harko14a}. In the rest of this paper,
we concentrate on the thermodynamic aspects
of the nonminimal torsion-matter coupling extension of teleparallel gravity.

\section{Non-equilibrium picture}
\label{sec:2}
 To study the thermodynamics of the nonminimal $f(T)$ gravity,
we rewrite the Eqs. (\ref{Friedmann1}) and (\ref{Friedmann2}) as follows
\begin{equation}
H^{2}=\frac{8\pi G}{3\mathcal{F}}(\hat{\rho}_{d}+\rho_{m}),\label{Friedmann 3}
\end{equation}
\begin{equation}
\dot{H}=-\frac{4\pi G}{\mathcal{F}}(\hat{\rho}_{d}+\hat{p}_{d}+\rho_{m}+p_{m}),\label{Friedmann 4}
\end{equation}
where the energy density and pressure of the dark components are defined as
\begin{equation}
\hat{\rho}_{d}\equiv\frac{1}{16\pi G}\left(TF_{1}-f_{1}\right)+\lambda\rho_{m}(f_{2}-TF_{2})\label{rho-dark}
 \,,
\end{equation}

\begin{equation}
\hat{p}_{d}\equiv\frac{1}{16\pi G}\big[f_{1}-TF_{1}+4H(\dot{F}_{1}-16\pi G\lambda\rho_{m}\dot{F}_{2})
+\lambda P_{m}(f_{2}-2TF_{2})-\lambda\rho_{m}TF_{2}\big]\label{p-dark}
\end{equation}
respectively. Here, a hat denotes quantities in the non-equilibrium description of thermodynamics
which do not satisfy the standard continuity equation so that
\begin{equation}
\dot{\hat{\rho}}_{d}+3H(\hat{\rho}_{d}+\hat{p}_{d})=\frac{1}{16\pi G}(-T\dot{F}_{1})+\lambda\rho_{m}T
\dot{F}_{2}-3\lambda THF_{2}(\rho_{m}+p_{m})\label{non-continiuty}\,.
\end{equation}
The perfect fluid satisfies the continuity equations by virtue of the Bianchi identity
\begin{equation}
\dot{\rho}_{m}+3H(\rho_{m}+p_{m})=0\label{continiuty-fluid}\,.
\end{equation}

\subsection{First law of thermodynamics}
\label{sec:3}
 Now we investigate the thermodynamic behavior of the nonminimal $f(T)$ gravity on the apparent horizon.
 In the flat FRW universe, the radius $\tilde{r}_{A}$ of the dynamical apparent horizon is given by \cite{akbar07}
\begin{equation}
\tilde{r}_{A}=\frac{1}{H},\label{radius}\
\end{equation}
 By taking the time derivative of this equation and substituting Eq. (\ref{Friedmann 4}) into
the result, we obtain
\begin{equation}
\frac{\mathcal{F}}{4\pi G}d\tilde{r}_{A}=H\tilde{r}^{3}_{A}(\hat{\rho}_{d}+\hat{p}_{d}+\rho_{m}+p_{m})dt\,.\label{dr}
\end{equation}
 In the Einstein gravity, the Bekenstein-Hawking relation $S = A/(4G)$ defines the horizon entropy,
where $A = 4\pi\tilde{r}^{2}_{A}$ is the area of the apparent horizon \cite{bardeen73}.
In the framework of the generalized theories of gravity such as $f(R)$ modified gravity, a horizon entropy $\hat{S}$
associated with a Noether charge, called the Wald entropy \cite{wald93}, is expressed
as $\hat{S}= A/(4G_{eff})$, where $G_{eff}$ is the effective gravitational coupling
\cite{brustein09}.

In the context of $f(T)$ gravity, it has been shown that the first law of black hole thermodynamics breaks down
 \cite{Miao11} due to the violation of local Lorentz invariance \cite{Li11}. However, it is argued that when
$f''=d^2f/dT^2$ is small, the entropy of the black hole in $f(T)$ gravity is approximately equal to $f'(T)A/4$.
furthermore, from the study of the matter density perturbations in $f(T)$ gravity, one can take
the effective gravitational coupling taken as $G_{eff}=G/f'(T)$.
Hence, similar to the $f(T)$ case, with the Friedmann equation (\ref{Friedmann 4}),
 we take the effective gravitational coupling as $G_{eff}=G/\mathcal{F}$,
so the Wald entropy in nonminimal $f(T)$ gravity is given by
\begin{equation}
\hat{S}=\frac{\mathcal{F}A}{4G}.\label{ Wald entropy }
\end{equation}

Using Eqs. (\ref{dr}) and (\ref{ Wald entropy }), we find
\begin{equation}
\frac{1}{2\pi \tilde{r}_{A}}d\hat{S}=4\pi \tilde{r}^{3}_{A}H\left(\hat{\rho}_{d}+\hat{p}_{d}+\rho_{m}+p_{m}\right)dt
+\frac{\tilde{r}_{A}}{2G}\dot{\mathcal{F}}dt.\label{entropy1}
\end{equation}
 The temperature of the apparent horizon is given by the Hawking temperature
$T_{h}=\frac{|\kappa_s|}{2\pi}$ where  $\kappa_s=-\frac{1}{{\tilde{r}}_{A}}(1-\frac{\dot{\tilde{r}}_
{A}}{2H\tilde{r}_{A}})$ is the surface gravity at the apparent horizon. Multiplying Eq. (\ref{entropy1})
with the term $1-\dot{\tilde{r}}_{A}/(2H\tilde{r}_{A})$ yields
\begin{equation}
T_{h}d\hat{S}=4\pi \tilde{r}^{3}_{A}H\left(\hat{\rho}_{d}+\hat{p}_{d}+\rho_{m}+p_{m}\right)dt
-2\pi\tilde{r}^{2}_{A}(\hat{\rho}_{d}+\hat{p}_{d}+\rho_{m}+p_{m})d\tilde{r}_{A}
+\frac{\pi\tilde{r}^{2}_{A}T_{h}}{G}d\mathcal{F},\label{entropy2}
\end{equation}
 In the Einstein gravity, the total energy inside a sphere of radius $\tilde{r}_{A }$
of the apparent horizon is $E=\tilde{r}_{A}/2G$. However, in the context of generalized gravity,
one should use the effective gravitational constant in this relation.
Hence, in the nonminimal modified $f(T)$ gravity, the total energy is given by
the following equation
\begin{equation}
\hat{E}=\frac{\tilde{r}_{A}\mathcal{F}}{2G}=\frac{3V\mathcal{F}}{8\pi G}
\left(H^2+\frac{K}{a^2}\right)=(\hat{\rho}_{d}+\rho_{m})V\,\label{energy1}
\end{equation}
 where $V = \frac{4}{3}\pi\tilde{r}^{3}_{A}$ is the volume of  3-dimensional sphere.
Taking the time derivative of Eq. (\ref{energy1}) we find
\begin{equation}
d\hat{E}=-4\pi \tilde{r}^{3}_{A}H(\hat{\rho}_{d}+\hat{p}_{d}+\rho_{m}
+p_{m})dt+4\pi\tilde{r}^{2}_{A}(\hat{\rho}_{d}+\rho_{m})d\tilde{r}_{A}
+\frac{\tilde{r}_{A}}{2G}d\mathcal{F}\,.\label{energy2}
\end{equation}
Using Eqs. (\ref{entropy2}) and (\ref{energy2}) leads to
\begin{equation}
T_{h}d\hat{S}=d\hat{E}+2\pi\tilde{r}^{2}_{A}(\hat{\rho}_{d}+
\rho_{m}-\hat{p}_{d}-p_{m})d\tilde{r}_{A}
+\frac{\tilde{r}_{A}}{2G}(1+2\pi\tilde{r}_{A}T_{h})d\mathcal{F}
\label{entropy3}\,.
\end{equation}
 By introducing the work density $\hat{W}=\frac{1}{2}(\hat{\rho}_{d}+\rho_{m}-\hat{p}_{d}-p_{m})$
\cite{hayward98}, one can rewritte the Eq.(\ref{entropy3}) as follows
\begin{equation}
T_{h}d\hat{S}=-d\hat{E}+\hat{W}dV+\frac{\tilde{r}_{A}}{2G}(1+2\pi\tilde{r}_{A}T_{h})d\mathcal{F}\,.
\end{equation}
 The above equation consists of additional term which is produced due to the non-equilibrium
representation of thermodynamics. Consequently, the first law of thermodynamics can be expressed as follows
\begin{equation}
T_{h}d\hat{S}+T_{h}d_{i}\hat{S}=-d\hat{E}+\hat{W}dV\,,\label{first law-non}
\end{equation}
where we have defined an entropy production term as
\begin{equation}
d_{i}\hat{S}=-\frac{\tilde{r}_{A}}{2GT_{h}}(1+2\pi\tilde{r}_{A}T_{h})d\mathcal{F}=
\frac{6\pi}{GT}\left(\frac{\dot{T}+8HT}{\dot{T}+4HT}\right)d\mathcal{F} \label{d_i hats}\,.
\end{equation}
 The appearance of the additional term $d_{i}\hat{S}$ illustrates that the horizon
thermodynamics is non-equilibrium one in the case of nonminimal $f(T)$ gravity. Indeed, the
violation of the standard first law of thermodynamics in this case is a result of the definition
of the dark energy momentum components as $\hat{\rho}_d$ and $\hat{p}_d$ which do not satisfy
the continuity equation (\ref{non-continiuty}).
In the next section we show that by definition of the energy density and pressure of this generalised $f(T)$ scenario in a way
 that the new components satisfy the continuity equation, it is possible to have an equilibrium description of thermodynamics so,
the first law of thermodynamics can be justified.

\subsection{Second law of thermodynamics}
\label{sec:4}
 In this subsection, we investigate the validity of the second law of thermodynamics
at the apparent horizon in the framework of the nonminimal torsion-matter coupling model.
 The second law of thermodynamics states that the sum
of the horizon entropy and the entropy of the matter field, i.e. the
total entropy, is a non-decreasing function of time. Assuming a same temperature between the outside and inside
of the apparent horizon, the condition to satisfy the second law of thermodynamics is given by
\begin{equation}
\frac{d\hat{S}}{dt} + \frac{d\left( d_i \hat{S} \right)}{dt}
+ \frac{d\hat{S}_{t}}{dt} \geq 0\,.
\label{SL1}
\end{equation}
 where $d\hat{S}$ and $d_i \hat{S}$ are deduced from the first law of
thermodynamics (Eq. \ref{first law-non}) and $d\hat{S}_{t}$ can be extracted from the
Gibb's equation which relates the entropy of all matter and energy sources to
the pressure in the horizon so that
\begin{equation}
T_{h} d\hat{S}_{t} = d\left( \hat{\rho}_{t} V \right) +\hat{p}_{t} dV
= V d\hat{\rho}_{t} + \left( \hat{\rho}_{t} + \hat{p}_{t} \right) dV\,,
\label{gibbs1}
\end{equation}
 where $T_{h}$ and $\hat{S}_{t}$ are corresponding to the temperature and
entropy of total energy inside the horizon,
respectively and we have defined $\hat{\rho}_{t}\equiv\rho_{m}
+\hat{\rho}_{d}$ and $\hat{p}_{t}\equiv p_{m}+\hat{p}_{d}$.
 Taking the time derivative of equation (\ref{gibbs1}) and using
(\ref{non-continiuty}) and (\ref{continiuty-fluid}), one can get
\begin{equation}
T_{h}\frac{d\hat{S}_{t}}{dt} =4\pi \tilde{r}^{2}_{A}(\hat{\rho}_{t}+
\hat{p}_{t})(\dot{\tilde{r}}_{A}-1)+\frac{\tilde{r}_{A}}{2G}\dot{\mathcal{F}}.
\label{gibbs2}
\end{equation}
 Now, substituting the above equation and Eq. (\ref{first law-non}) in (\ref{SL1}), we find
\begin{equation}
\frac{1}{2G} \frac{\dot{H}^2 \mathcal{F}}{H^4}\geq 0\,.
\label{SL2}
\end{equation}
 This result describes the validity of the second law of thermodynamics in the non-equilibrium
treatment. So the condition needed to hold the second law of thermodynamics in nonminimal $f(T)$ gravity is
equivalent to $\mathcal{F}\geq0$. Note that $\mathcal{F}$ should be positive in order to $\hat{E}\geq0$.
This condition imposes a constraint to the coupling parameter $\lambda$, so that
$\lambda F_{2}<\frac{1+2F_{1}}{32\pi G\rho_{m}}$. As a result, the upper bound of lambda depends explicitly to the
choices of two functions $f_1(T)$ and $f_2(T)$.
In the flat FRW universe, the effective equation of state parameter is defined as \cite{Nojiri07}
\begin{equation}
\omega_{eff}=-\left(1+\frac{2\dot{H}}{3H^2}\right).
\label{omega}
\end{equation}
 Here $\omega_{eff}>-1$, $\dot{H}<0$, represents the quintessence phase of the
universe while $\omega_{eff}>-1$ , $\dot{H}>0$, is corresponding to the phantom phase.
From equation (\ref{SL1}) we find that in the non-equilibrium picture, the second law of
thermodynamics is satisfied in both phantom and quintessence phases
of the universe evolution.

\section{Equilibrium picture}

 In this section we investigate the possibility to have an equilibrium picture of thermodynamics
in the nonminimal $f(T)$ modified gravity setup.
\label{sec:6}
 To do this, we rewrite the Friedmann equations
(\ref{Friedmann1}) and (\ref{Friedmann2}) in the following form
\begin{equation}
H^{2}=\frac{8\pi G}{3}\left[1+\lambda(f_{2}-2TF_{2})\rho_{m}\right]-\frac{1}{6}(f_{1}-2TF_{1})\label{H2-equi}
\end{equation}
and
\begin{equation}
\dot{H}=-\frac{8\pi G(\rho_{m}+p_{m})\left[1+\lambda(f_{2}-2T
F_{2})\right]}{1+\mathcal{F}+4T(F'_{1}-16\pi G\lambda\rho_{m}F'_{2})}\label{dot_H-equi}
\end{equation}
 Equations (\ref{H2-equi}) and (\ref{dot_H-equi}) can be expressed as
\begin{equation}
3H^{2}=8\pi G(\rho_{d}+\rho_{m})
\end{equation}
and
\begin{equation}
2\dot{H}+3H^{2}=-8\pi G(p_{d}+p_{m})
\end{equation}
where we have defined the energy density and pressure of dark components as
\begin{equation}
\rho_{d}\equiv\lambda\rho_{m}f_{2}-\frac{1}{16\pi G}[f_{1}+T(1-\mathcal{F})]\,,
\end{equation}
and
\begin{equation}
p_{d}\equiv(\rho_{m}+p_{m})\left[\frac{2+2\lambda(f_{2}-2T
F_{2})}{1+\mathcal{F}+4T(F'_{1}-16\pi G\lambda\rho_{m}F'_{2})}-1\right]
-\lambda\rho_{m}f_{2}+\frac{1}{16\pi G}\big[f_{1}+T(1-\mathcal{F})\big]
\end{equation}
respectively. Now, from the new definition of the dark energy energy density and pressure,
the standard continuity equation can be retrieved as follows
\begin{equation}
\dot{\rho_{d}}+3H(\rho_{d}+p_{d})=0 .
\end{equation}
 So, the equilibrium description of thermodynamics can be treated in a same manner as general relativity.
\subsection{First law of thermodynamics}
In the new representation of the dark energy components, the time derivative of the dynamically apparent horizon is given by
\begin{equation}
\frac{d\tilde{r}_{A}}{dt}=4\pi G\tilde{r}^{3}_{A}H(\rho_{d}+p_{d}+\rho_{m}+p_{m})\,.
\end{equation}
Introducing the Bekenstein-Hawking entropy as $S=A/(4G)$, we find
\begin{equation}
\frac{1}{2\pi\tilde{r}_{A}}dS=4\pi\tilde{r}_{A}^{3}H(\rho_{d}+p_{d}+\rho_{m}+p_{m})dt.\label{dre}
\end{equation}
Substituting the horizon temperature ($T_{h}=-\frac{1}{2\pi{\tilde{r}}_{A}}[1-\frac{\dot{\tilde{r}}_
{A}}{2H\tilde{r}_{A}}]$ ) in the above equation leads to
\begin{equation}
T_{h}dS=4\pi\tilde{r}_{A}^{3}H(\rho_{d}+p_{d}+\rho_{m}+p_{m})dt
-2\pi\tilde{r}_{A}^{2}
(\rho_{d}+p_{d}+\rho_{m}+p_{m})d\tilde{r}_{A}\label{tds}\,.
\end{equation}
 Defining the Misner-Sharp energy as $E=\tilde{r}_{A}/(2G)=V(\rho_{d}+\rho_{m})$ in the Einstein gravity, we get
\begin{equation}
dE=-4\pi\tilde{r}_{A}^{3}H(\rho_{d}+p_{d}+\rho_{m}+p_{m})dt+
4\pi\tilde{r}_{A}^{2}(\rho_{d}+\rho_{m})d\tilde{r}_{A}
\label{misner}\,.
\end{equation}
With the definition of the work density as $W=(\rho_{d}+\rho_{m}-p_{d}-p_{m})/2$ \cite{hayward98} and using Eqs. (\ref{tds}) and
(\ref{misner}), we obtain the following equation corresponding to the first law of equilibrium thermodynamics
\begin{equation}
T_{h}dS=-dE+WdV\label{first low}\,.
\end{equation}
 As a result, an equilibrium treatment of the first law of thermodynamics is obtained with a suitable definition for
the energy momentum tensor components of dark energy.

One can show that the horizon entropy in the equilibrium picture has a relation with the horizon entropy in
the non-equilibrium picture as
\begin{equation}
dS=d\hat{S}+d_{i}\hat{S}+\frac{\tilde{r}_{A}}{2GT_{h}}d\mathcal{F}-\frac{2\pi\dot{H}(1-\mathcal{F})}{GH^3}dt.
\label{ds-di_s1}
\end{equation}
Using equations (\ref{d_i hats}) and (\ref{dre}), the above relation reduces to the following form
\begin{equation}
dS=\frac{1}{\mathcal{F}}\left(d\hat{S}+\frac{\dot{H}+2H^2}{\dot{H}+2H^2}d_{i}\hat{S}\right).
\label{ds-di_s2}
\end{equation}
The difference between $S$ and $\hat{S}$  appears in the nonminimal $f(T)$ gravity
is due to $d\mathcal{F}/dT=0$. Note that in the pure teleparallel gravity case,
which corresponds to $f_{1}(T)=T$ and $\lambda=0$,
 we have $\hat{S}=S$.
  From Eq. (\ref{ds-di_s2}), we see that the change of the entropy $S$
  in the equilibrium picture involves the information of both $d\hat{S}$
  and $d_{i}\hat{S}$ in the non-equilibrium picture.
\subsection{Second law of thermodynamics}
 In this subsection we use the first law of thermodynamics in the
 equilibrium picture (Eq. \ref{first low}) to find the general condition
which is needed to hold the second law of thermodynamics in the nonminimal $f(T)$ scenario.
As before, we assume a same temperature for outside and inside the apparent horizon of the universe.
To examine the validity of the second law of thermodynamics,
we consider the Gibbs relation in terms of all matter and energy components
\begin{equation}
T_{h} dS_{t}=Vd\rho_{t}+\left(\rho_{t}+p_{t}\right)dV\,.
\label{gibs2}
\end{equation}
where  $\rho_{t}\equiv\rho_{m}
+\rho_{d}$ and $p_{t}\equiv p_{m}+p_{d}$.
 Using equations (\ref{dot_H-equi})and (\ref{tds}), the evolution of the horizon entropy is given by
\begin{equation}
\dot{S} = 8\pi^2 H \tilde{r}_A^4 \left(\rho_{t}+p_{t}\right)
=\frac{6\pi}{G} \frac{\dot{T}}{T^2}\,.
\label{dots2}
\end{equation}
 To obey the second law of thermodynamics in the equilibrium picture of nonminimal $f(T)$ gravity, we
require that $\frac{dS}{dt}+\frac{dS_{t}}{dt}\geq 0$. Plugging Eqs. (\ref{gibs2}) and (\ref{dots2})
into this condition, we find
\begin{equation}
-\left(4HT +\dot{T} \right)=12H\left(2H^2+\dot{H}\right)
\geq 0\,.
\label{2law}
\end{equation}
 It is clear that this condition naturally holds in the phantom phase of the universe which
$\dot{H}>0$. On the other hand, for the validity of the second law in the quintessence phase ($\dot{H}<0$)
it is required that $\dot{H}<2H^2$. Note that the result (\ref{2law}) is independent of the forms of two
functions $f_i(T)$ and imposes no constraint to the coupling parameter $\lambda$.

\section{CONCLUSIONS}
\label{conclu}
 In this paper, we have studied the laws of thermodynamics in a generalized teleparallel
gravity which contains a nonminimal coupling between the matter field
and an arbitrary function of the torsion scalar.
 From the Friedmann equations, we have constructed
the first law of thermodynamics on the apparent horizon in two different approaches.
These approaches are depend on
the definition of the components of the energy momentum tensor of dark energy and lead to the
non-equilibrium and equilibrium descriptions of thermodynamics.
We have seen that an entropy production term appears in the non-equilibrium description due to
the violation of the standard continuity equation because of the definition of the dark energy components.
Consequently, the usual first law of thermodynamics is violated in the non-equilibrium picture.
 We also have examined the second law of thermodynamics which illustrates
that the total entropy evolution with time including the horizon entropy, the non-equilibrium
entropy production term, and the entropy of all matter field and
energy components is a non-decreasing function of time.
We have found that the second law of thermodynamics is satisfied in both phantom and quintessence phases
of the universe evolution.
It should be noted that the validity of the second law of thermodynamics in the non-equilibrium picture
imposes a constraint on the
coupling parameter $\lambda$, i.e. an upper bound on $\lambda$ can be deduced which depends explicitly to the
choices of functions $f_1(T)$ and $f_2(T)$.

 To have an equilibrium description of thermodynamics, we have redefined the dark energy density and pressure
in a manner that the new components satisfy the continuity equation so, there is no extra entropy terms and the
first law of thermodynamics holds. We have also show that that the change of the horizon entropy
in the equilibrium picture involves the information of the change of the horizon entropy as well as the change of the
entropy production term in the non-equilibrium description.

In studying the second law of thermodynamics,
 we have examined the evolution of the entropy contributed by the horizon entropy and all matter fields and
energy contents in the equilibrium description. We have found the condition to satisfy the second law of thermodynamics
which holds in the phantom phase of the universe. Furthermore, for the validity of the second law in the non-phantom phase
it is required that $\dot{H}<2H^2$ and there is no constraint to the coupling parameter of the nonminimal $f(T)$ gravity scenario.
\section{Acknowledgements}
We would like to acknowledge Prof. Kourosh Nozari for invaluable remarks.

\end{document}